\documentstyle[12pt]{article}
\begin{document}
\begin{titlepage}
\begin{flushright}
TU-496\\
1996\\
\end{flushright}

\begin{center}
\LARGE{High Momentum Behavior of\\ Geometric Bremsstrahlung\\
 in the Expanding Universe }
\end{center}
\ \\
\ \\
\ \\
\begin{center}
{\Large{M.Hotta,\ H.\ Inoue,\  I.\ Joichi and M.\ Tanaka}}\\
\it{Department of Physics, Faculty of Science, Tohoku University,
 Sendai, 980-77, Japan}

\end{center}

\vspace{3cm}

\begin{abstract}
 We shall discuss various kinds of geometric bremsstrahlung processes 
in the spatially flat Robertson-Walker universe. Despite that the 
temperature of the universe is much higher than particle masses and 
the Hubble parameter, the transition probability of these processes 
do not vanish. It is also pointed out that explicit forms of the 
probability possess a new duality with respect to scale factor of 
background geometry.  
\end{abstract}

\end{titlepage}

%%%%%%%%%%%%%%%%%%%%%%%%%%%%%%%%%%%%%%%%%%%%%%%%%%%%%%%%%%%%%%%%%

\begin{flushleft}
\bf{1.\ \ 
Introduction
}
\end{flushleft}

Particles in the early universe    
undergo severe redshift as a result of the cosmic expansion 
and get decelerated extraordinarily in the comoving coordinates. 
 Consequently  radiation or massless particles may be emitted from the 
decelerated particle. This process induced by the geometry of the universe
 is regarded as a kind of bremsstrahlung. Thus we call it 
geometric bremsstrahlung. This effect may bring about
 many sorts of decay and emission process, which are 
 prohibited kinematically in the flat spacetime.

Despite of its ease to recognize  such a   
process in the classical mechanical meaning,  
there is a nontrivial aspect on  existence of 
 quantum geometric bremsstrahlung. 
 Temperature in many interesting situations of the early universe 
 is  much higher than
 particle masses and  the Hubble parameter. 
 Since momentum of the particle is comparable with the temperature, 
one might naively think that  we can neglect all the mass parameters 
 even in calculation of the transition probabilities.
 Hence it might be expected that
 the probability of the process is 
 equal to that calculated in massless theories in the flat spacetime, 
thus exactly zero. However this naive expectation is turned out to be
 wrong by  careful analysis.

In \cite{FHIY}, the first precise analysis has been performed 
in the four dimensional Robertson-Walker universe.
 They shows that  even in a high momentum limit there remains  
 the nonvanishing probability of 
photon emission via the geometric bremsstrahlung.

In this paper we shall give an extended analysis on several kinds of 
processes of the geometric bremsstrahlung. 
It includes analyses  on  the $\phi^3$ theory 
in arbitrary dimensional spacetimes,  the theory with Yukawa
interaction and  the massive vector field theory.
It will be shown that the high momentum limit, or the high temperature
limit, does not terminate the geometric bremsstrahlung process 
in such a rather wide class of interactions and in arbitrary dimension.
It is also  stressed that a new type of duality can be found 
in the forms of the transition probabilities for renormalizable
interactions.

In section 2 we introduce the spatially flat Robertson-Walker universe with
 past and future asymptotic flat regions and explain why we especially 
 consider it. In section 3 we shall give a review on free fields in 
our model of the universe. In section 4  
the geometric bremsstrahlung  induced by the Yukawa interaction
 is analyzed in detail. In section 5  the $\phi^3$ theory in 
 arbitrary dimensional spacetimes is surveyed. In section 6
 we give explicit forms of transition probability of the process  
including a massive vector field.

%%%%%%%%%%%%%%%%%%%%%%%%%%%%%%%%%%%%%%%%%%%%%%%%%%%%%%%%%%%%%%%%%%%%%%

\begin{flushleft}
\bf{2.\ \  Robertson-Walker Universe\\ \ \ \ \ 
 with Past and Future Minkowskian Regions }
\end{flushleft}

Here let us comment a little bit on our model of the Robertson-Walker 
universe.  In this paper only the spatially
flat model is argued, but we believe that our results will be extended
 to open and closed models. 
 The geometry of the spacetime  
is expressed by a metric tensor whose form is written as
$
g_{\mu\nu} =a(t)^2 \eta_{\mu\nu},
$
where $a(t)$ is a scale factor function and $\eta_{\mu\nu}$
 is the Minkowskian metric. The argument of the scale factor, $t$,  
is conformal time 
and proper time $\tau$ can be defined by $d\tau=a(t)dt$. From the 
metric the Christoffel symbol is easily manipulated as follows.
\begin{eqnarray}
\Gamma^\alpha_{\beta\gamma}
=\delta^\alpha_\beta \partial_\gamma \ln a
+\delta^\alpha_\gamma \partial_\beta \ln a
-\eta_{\beta\gamma} \partial^\alpha \ln a,
\end{eqnarray}
where $\partial^\alpha =\eta^{\alpha\beta} \partial_\beta$.
This yields an explicit form of the scalar curvature
\begin{eqnarray}
R=-\frac{1}{a^2} 
\left[ 2(n-1) \frac{d^2}{dt^2} \ln a
+(n-1)(n-2) \left(\frac{d}{dt} \ln a\right)^2
\right],
\end{eqnarray}
where $n$ is the dimension of the spacetime. 
 The Hubble parameter 
 is defined as usual by the scale factor as 
$$
H(t)=\frac{1}{a}\frac{da}{d\tau}=\frac{1}{a^2}\frac{da}{dt}.
$$

In later sections we must take account of interaction 
in the expanding universe. Then well-defined asymptotic free field is
required to construct S matrix elements rigorously. 
 However in general spacetimes 
 the construction of asymptotic fields often encounters some difficulties
\cite{BD}. 

 For example the universe accompanied with the big-bang 
possesses a definite birth time and an initial singularity. 
 Near at the birth time it needs more pieces of information (maybe quantum
gravity) to specify how to define asymptotic in-field. 

  The simplest prescription to avoid such problems is to 
  restrict ourselves on analysis in the universe 
equipped with Minkowskian past and future regions. 
 Clearly this enables us to define both asymptotic in- and out- fields
 just like those in the flat spacetime. It should be emphasized that 
 physically meaningful results derived from this model  
must be independent of artificial detail of the model, 
that is, how the universe gets into 
 and out of the expanding era. It will be found that results obtained
later satisfy this criterion.

Owing to appearance of the asymptotic flat regions the scale factor
must satisfy two constraints.
By virtue of the time rescaling invariance 
 one of them is expressed, losing no generality, as
$$a(t\sim \infty) =1 .$$
Now let $b$ denote ratio of initial scale factor to final one. Then 
another constraint is written as 
$$
a(t\sim -\infty)=b. 
$$

It is worth grasping qualitative behavior of Fourier components
of $a(t)^n$,
$$
F_n (\omega)=\int^{\infty}_{-\infty} dt a(t)^n e^{i\omega t} .
$$
Now let $\omega^{(n)}_{min} <\omega^{(n)}_1<\omega^{(n)}_2< \cdots
<\omega^{(n)}_{max}$ denote typical frequencies characterizing detailed
evolution  of
$a(t)^n$ with the following condition satisfied.
$$
F_n (\omega^{(n)}_i ) \sim O(1).
$$
It is natural to think that  all the $\omega^{(n)}_i$ are of 
order of the Hubble parameter in the expansion era. 

 For  $\omega\ll \omega^{(n)}_{min}$ 
it can be shown  that
\begin{eqnarray}
F_n (\omega \sim 0) \sim \frac{i}{\omega} (1-b^n ). \label{22.1}
\end{eqnarray}
Note that 
 this relation (\ref{22.1}) holds for rather arbitrary cosmic evolution.

On the other hand for  $\omega \gg \omega^{(n)}_{max}$ 
 $F_n$  behaves as  
$$
F_n (\omega\sim\infty) \sim 0.
$$

\pagebreak

%%%%%%%%%%%%%%%%%%%%%%%%%%%%%%%%%%%%%%%%%%%%%%%%%%%%%%%%%%%%%%%%%%%%%%%

\begin{flushleft}
\bf{3.\ \ Free Fields in the  Robertson-Walker Universe  }
\end{flushleft}

In this section 
 free fields in the universe introduced in section 2 are reviewed for later 
convenience.

Let us first review free scalar field in the expanding universe. 
The action with the
conformal coupling term is written as 
\begin{eqnarray}
S_{scalar} =\int d^n x \sqrt{|g|}
\left[\ 
\frac{1}{2}\left(\nabla \phi \right)^2
-\frac{1}{2}
\left( 
m^2 - \frac{n-2}{4(n-1)} R
\right)
\phi^2\ 
\right] .
\end{eqnarray}
Next let us change the field variable $\phi$ into 
 $\tilde{\phi}= a(t)^{n/2-1} \phi$ . Then the action is reduced into
\begin{eqnarray}
S_{scalar} =\int d^n x
\left[\ 
\frac{1}{2}\left(\partial \tilde{\phi} \right)^2
-\frac{1}{2}
\left( m a(t) \right)^2 \tilde{\phi}^2 \ 
\right] . \label{22.11}
\end{eqnarray}
This is just a free field action with time dependent mass $ma(t)$
in the flat spacetime.

Boundary conditions 
 for  asymptotic fields can be described 
 in terms of the rescaled field $\tilde{\phi}$. 
It is worthwhile to point out 
that spatially flat spacetimes possess isometry 
in the spatial section. Consequently
 the Fourier transformation  can be used in 
 equation of motion derived from eqn $(\ref{22.11})$. Therefore 
what we need is a solution which form is such that
\begin{eqnarray}
\tilde{\phi}(t,\vec{x}) = u_{\vec{p}} (t) e^{i\vec{p}\cdot\vec{x}},
\end{eqnarray}
where $\vec{p}$ is conserved conformal momentum. Its related 
physical momentum is expressed as $\vec{p}_{phys} =\vec{p}/ a(t)$. 
 It is  easily shown that this $u_{\vec{p}}$ satisfies 
a Schr{\"o}dinger-type equation,
\begin{eqnarray}
\left[
-\frac{d^2}{dt^2} -m^2 a(t)^2
\right]
u_{\vec{p}} = p^2 u_{\vec{p}}, \label{2.2}
\end{eqnarray}
where $p=|\vec{p}|$.
The in-mode function of this equation is specified with the following
 boundary condition,
\begin{eqnarray}
u^{in}_{\vec{p}} (t\sim-\infty)=
 \frac{1}{\sqrt{(2\pi)^{n-1} 2\sqrt{p^2 +m^2 b^2} }}
 e^{- i  t\sqrt{p^2 + m^2 b^2}} .
\end{eqnarray}
On the other hand,  the out-mode function  satisfies another
 boundary condition,
\begin{eqnarray}
u^{out}_{\vec{p}} (t\sim\infty)=
 \frac{1}{\sqrt{(2\pi)^{n-1} 2\sqrt{p^2 +m^2 } }}
 e^{- i  t\sqrt{p^2 + m^2 }} .
\end{eqnarray}
The in-(out-) mode function may have reflection wave terms in $t\rightarrow
  \infty (t\rightarrow -\infty)$ 
 induced by the nontrivial potential term, $-m^2 a(t)^2$. 
Usually such  existence of the reflection wave means particle 
creation from the vacuum state in the field theoretical context. 
 Typical energy of the created particle
 is of order of the typical Hubble parameter. 
Hence the energy can be thought much smaller 
than temperature of the universe.  
This effect is of some interest, however  
 it is not our target in this report. In fact 
 we shall concentrate our attention on 
particles with high momentum nearly equal to the temperature.  
Consequently the reflection wave term is negligible in the following 
 argument.

Now we  can also exhibit 
 results for a soluble example 
with  a step scale factor  
$a(t) = b\Theta (-t) +\Theta(t)$. 
For example analytic form of the in-mode function  is expressed as follows.
\begin{eqnarray}
u^{in}_{\vec{p}} (t<0) &=& \frac{1}{\sqrt{(2\pi)^{n-1}
 2\sqrt{p^2 + m^2 b^2}}}
e^{-it\sqrt{p^2 +m^2 b^2} },\nonumber\\
u^{in}_{\vec{p}} (t>0) &=& \frac{1}{\sqrt{(2\pi)^{n-1} 2\sqrt{p^2 + m^2 b^2 }}}
\left[
A e^{-it\sqrt{p^2 +m^2 } }+B e^{it\sqrt{p^2 +m^2 } }
\right],\nonumber
\end{eqnarray}
where
\begin{eqnarray}
A&=&\frac{1}{2}
\left(
1+\sqrt{\frac{p^2 +m^2 b^2 }{p^2 + m^2}}
\right),\\
B&=&\frac{1}{2}\left(
1-\sqrt{\frac{p^2 +m^2 b^2 }{p^2 + m^2}}
\right).\label{2.1}
\end{eqnarray}
This will be used in section 4.
 As it should be, the reflection coefficient $B$ vanishes in 
$p\rightarrow\infty$. 

Here we also comment on the WKB amplitudes of eqn(\ref{2.2}). 
When a high momentum condition;
 $p^3 \gg m^2 a\frac{da}{dt}$ or $p_{phys}^3 \gg m^2 H$ 
is satisfied, the WKB approximation is validated enough 
in eqn(\ref{2.2}). 
Then reflection coefficient can be neglected and both 
in- and out-mode functions are written in the same form as follows.
\begin{eqnarray}
u_{\vec{p}} \sim \frac{1}{\sqrt{(2\pi)^3 2E(p,t)}} e^{- i \int dt E(p,t)},
\end{eqnarray}
where $E(p,t)= \sqrt{p^2 +m^2 a(t)^2 }$. 
In the early universe situation, this form  gives us 
 reliable estimation of the wavefunction.

\vspace{1cm}

Next let us review free propagation of spinor field. 
Writing down the action needs vielbein $e^a_\mu$  
 related with the metric tensor like $g_{\mu\nu} = e^a_{\mu} e_{a\nu}$.
For the spatially flat Robertson-Walker universe the vielbein reads 
\begin{eqnarray}
e^a_{\mu}= a(t) \delta^a_\mu,\ \ \ e^{\mu}_a =a(t)^{-1} \delta^\mu_a .
\end{eqnarray}
Then spin connection is obtained from this  vielbein such that
\begin{eqnarray}
\omega^{ab}_{\mu}= e^a_{\lambda} \nabla_\mu e^{b\lambda}
= \delta^a_\mu \partial^b \ln a
-\delta^b_\mu \partial^a \ln a.
\end{eqnarray}

The action of free spinor field reads 
\begin{eqnarray}
S_{spinor} =\int d^n x \det [e^a_\mu]
\left[
\bar{\Psi} i \gamma^\mu \nabla_\mu \Psi
-m \bar{\Psi}\Psi
\right],
\end{eqnarray}
where
\begin{eqnarray}
&&\gamma^\mu = e^\mu_a \gamma^a,\\
&&\{\gamma_a , \gamma_b \} =2\eta_{ab},\\
&&\nabla_\mu \Psi = 
\left(
\partial_\mu - \frac{i}{4}\omega^{ab}_\mu \sigma_{ab}
\right)
\Psi,\\
&&\sigma_{ab} =\frac{i}{2} [\gamma_a ,\gamma_b ].
\end{eqnarray}
Subsequently
 by defining a rescale field $\tilde{\Psi}= a^{\frac{n-1}{2}} \Psi$
 we rewrite the action as  
\begin{eqnarray}
S_{spinor} =\int d^n x 
\left[
\bar{\tilde{\Psi}} i \gamma^a \partial_a \tilde{\Psi}
-ma(t) \bar{\tilde{\Psi}}\tilde{\Psi}
\right]
\end{eqnarray}
where we have used the relations;
$\gamma^a \gamma_a =n$ and $\gamma^a \gamma^b \gamma_a =(2-n)\gamma^b$.
Thus, like the scalar field, 
the theory can be also reduced into just a free 
 theory with time dependent mass in the flat spacetimes.

Next 
 let us introduce  mode functions more explicitly for $n=4$. Here  
 we adopt the standard representation for the gamma matrices,
\begin{eqnarray}
\gamma^0 =
\left[
\begin{array}{cc}
\bf{1} & \bf{0}\\
\bf{0} & -\bf{1} 
\end{array}
\right]
,\ \ 
\vec{\gamma} =
\left[
\begin{array}{cc}
\bf{0} & \vec{\sigma}\\
-\vec{\sigma} & \bf{0} 
\end{array}
\right].
\end{eqnarray}
Also we introduce 
$\alpha_{h,\vec{p}}$ and $\beta_{h,\vec{p}}$
defined by the following equations,
\begin{eqnarray}
\tilde{\Psi}=e^{i\vec{p}\cdot\vec{x}}
\left[
\begin{array}{c}
\alpha_{h,\vec{p}} (t) \xi(h,\vec{p})\\
\beta_{h,\vec{p}} (t) \xi(h,\vec{p})
\end{array}
\right],\ \ 
\vec{\sigma}\cdot\vec{p} \xi(h,\vec{p}) =h|\vec{p}| \xi(h, \vec{p}),
\end{eqnarray}
where $h= \pm 1$ and $h/2$ is helicity of the particle.
Then $\alpha_{h,\vec{p}}$ and $\beta_{h,\vec{p}}$ satisfy
equations such that
\begin{eqnarray}
&&\beta_{h,\vec{p}} (t) =\frac{1}{h|\vec{p}|}
\left[
i\frac{d}{dt} -ma(t)
\right]
\alpha_{h,\vec{p}} (t) ,\\
&&
\left[
\frac{d^2}{dt^2} +|\vec{p}|^2 + m^2 a(t)^2 +im \frac{d}{dt} a
\right]
\alpha_{h,\vec{p}} (t) =0 .
\end{eqnarray}
The in- and out- mode functions for 
$\alpha_{h,\vec{p}}$ are specified by imposing boundary conditions,
\begin{eqnarray}
\alpha^{in(\pm)}_{h,\vec{p}}(t\sim-\infty)
=\frac{\sqrt{\sqrt{p^2+m^2 b^2}\pm mb }}
{\sqrt{(2\pi)^3 2\sqrt{p^2 +m^2 b^2} } }
e^{\mp i t \sqrt{p^2 +m^2 b^2}  },
\end{eqnarray}
\begin{eqnarray}
\alpha^{out(\pm)}_{h,\vec{p}}(t\sim\infty)
=\frac{\sqrt{\sqrt{p^2+m^2 }\pm m }}
{\sqrt{(2\pi)^3 2\sqrt{p^2 +m^2 } } }
e^{\mp i t \sqrt{p^2 +m^2 }  } ,
\end{eqnarray}
where the sign $+( - )$ in $\alpha^{(\pm)}$ 
corresponds to particle(antiparticle) wavefunction.

For a step evolution like $a(t) = b\Theta (-t) +\Theta (t)$, exact
analysis 
is possible and for example analytic 
 $in(+)$-mode function is given by
\begin{eqnarray}
\alpha^{in(+)}_{h, \vec{p}} (t<0) 
&=& \frac{\sqrt{\sqrt{p^2 + m^2 b^2} +mb}}{\sqrt{(2\pi)^3
 2\sqrt{p^2 + m^2 b^2}}}
e^{-it\sqrt{p^2 +m^2 b^2} }\nonumber\\
\alpha^{in(+)}_{h,\vec{p}} (t>0) &=& 
\frac{\sqrt{\sqrt{p^2 + m^2 b^2} +mb}}{\sqrt{(2\pi)^3 2\sqrt{p^2 + m^2 b^2}}}
\left[
A_f e^{-it\sqrt{p^2 +m^2 } }
+B_f e^{it\sqrt{p^2 +m^2 } }
\right]\nonumber
\end{eqnarray}
where
\begin{eqnarray}
A_f &=&\frac{1}{2}
\left(
1+\sqrt{\frac{p^2 +m^2 b^2 }{p^2 + m^2}}
+\frac{m(1-b)}{\sqrt{p^2 + m^2}}
\right),\\
B_f &=&\frac{1}{2}\left(
1-\sqrt{\frac{p^2 +m^2 b^2 }{p^2 + m^2}}
-\frac{m(1-b)}{\sqrt{p^2 + m^2}}
\right).
\end{eqnarray}

For the spinor field the WKB approximation also 
 can be justified when $p \gg m a$ or $p_{phys}\gg m$.
Then the following amplitude is obtained. 
\begin{eqnarray}
\alpha^{(\pm)}_{h, \vec{p}} (t)\sim \frac{\sqrt{p\pm ma}}{\sqrt{(2\pi)^3 2p}}
e^{\mp i (pt +\frac{m^2}{2p}\int dt a(t)^2) } . 
\end{eqnarray}

It will be also useful in the later sections to introduce  $U$ and $V$
spinors corresponding to particle and antiparticle as follows.
\begin{eqnarray}
U(h,\vec{p}, a)=
\left[
\begin{array}{c}
 \sqrt{E(p, a(t))+ma}\xi(h,\vec{p})\\
h \sqrt{E(p,a(t))-ma} \xi(h,\vec{p})
\end{array}
\right],
\ \ \ 
V(h,\vec{p}, a)=
\left[
\begin{array}{c}
- h\sqrt{E(p,a(t))-ma}\eta(h,\vec{p})\\
 \sqrt{E(p,a(t))+ma} \eta(h,\vec{p})
\end{array}
\right]\nonumber
\end{eqnarray}
where $E(\vec{p}, a(t))=\sqrt{p^2 +m^2 a(t)^2}$ and 
$\eta(h, \vec{p}) = -i\sigma^2 \xi^{\ast}(h,\vec{p})$.
Using a polar parametrization such that
\begin{eqnarray}
\vec{p}=p(\sin \theta \cos \phi , \sin \theta \sin \phi , \cos \theta ),
\end{eqnarray}
the explicit forms of $\xi$ and $\eta$ are written down as follows.
\begin{eqnarray}
\xi(1,\vec{p})=
\left[
\begin{array}{c}
e^{-i\phi/2} 
\cos \left(\frac{\theta}{2}\right) \\
e^{i\phi/2} 
\sin \left(\frac{\theta}{2}\right) 
\end{array}
\right],
\ \ \ \ 
\xi(-1,\vec{p})=
\left[
\begin{array}{c}
-e^{-i\phi/2} 
\sin \left(\frac{\theta}{2}\right) \\
e^{i\phi/2} 
\cos \left(\frac{\theta}{2}\right) 
\end{array}
\right],
\end{eqnarray}

\begin{eqnarray}
\eta(1,\vec{p})=
\left[
\begin{array}{c}
-e^{-i\phi/2} 
\sin \left(\frac{\theta}{2}\right) \\
e^{i\phi/2} 
\cos \left(\frac{\theta}{2}\right) 
\end{array}
\right],
\ \ \ \ 
\eta(-1,\vec{p})=
\left[
\begin{array}{c}
-e^{-i\phi/2} 
\cos \left(\frac{\theta}{2}\right) \\
-e^{i\phi/2} 
\sin \left(\frac{\theta}{2}\right) 
\end{array}
\right].
\end{eqnarray}

\vspace{1cm}

Next  let us give a review of  massive vector field in the four dimension. 
The original equation of motion is written as follows.
\begin{eqnarray}
\nabla^\mu F_{\mu\nu} +m^2 A_\nu =0.
\end{eqnarray}
where $F_{\mu\nu} = \nabla_\mu A_\nu -\nabla_\nu A_\mu
=\partial_\mu A_\nu -\partial_\nu A_\mu$ .\\
Now using the conformal flatness $g_{\mu\nu} =a(t)^2
\eta_{\mu\nu}$, the equation is rewritten as 
\begin{eqnarray}
\left[
\partial^2 + m^2 a^2
\right]A_\mu 
-\partial_\mu (\partial A)
=0.
\end{eqnarray}
The transverse wave solution can be introduced such that
\begin{eqnarray}
\left[ A^{(h)}_\mu \right] =e^{i\vec{p}\cdot\vec{x}} 
\tilde{u}^{(h)}_{\vec{p}} (t)
\left[
\begin{array}{c}
0 \\
-\vec{n}^{(h)} 
\end{array}
\right],
\end{eqnarray}
where $h=\pm 1$ and  $\vec{n}^{(h)}$ is a unit vector satisfying 
 $\vec{p}\cdot\vec{n}^{(h)} =0$. The equation of motion requires 
\begin{eqnarray}
\left[
\frac{d^2}{dt^2} +p^2 +m^2 a(t)^2
\right]
\tilde{u}^{(h)}_{\vec{p}}  (t) =0 .\nonumber
\end{eqnarray}
On the other hand the form of 
 longitudinal wave solution is 
\begin{eqnarray}
\left[ A^{(L)}_\mu \right]
=
e^{i\vec{p}\cdot\vec{x}} 
\left[
\begin{array}{c}
\alpha^{(L)}(t) \\
-\frac{\vec{p}}{p} \beta^{(L)}(t) 
\end{array}
\right].
\end{eqnarray}
Time dependent factors $\alpha^{(L)}$ and $\beta^{(L)}$ need to satisfy
\begin{eqnarray}
\alpha^{(L)}=\frac{i p}{p^2 +m^2 a(t)^2 }\frac{d\beta^{(L)}}{dt}.
\end{eqnarray}
Rescaling such that 
\begin{eqnarray}
\beta^{(L)} =
\sqrt{\frac{p^2 +m^2 a^2}{m^2 a^2} }\tilde{\beta},
\end{eqnarray}
 the following  Schr{\"o}dinger type equation also should hold, 
\begin{eqnarray}
\left[
\frac{d^2}{dt^2}+p^2 + m^2 a^2
+
\frac{3m^2 p^2}{(p^2 + m^2 a^2)^2 } \left(\frac{da}{dt}\right)^2
-
\frac{p^2}{p^2 +m^2 a^2} \frac{1}{a}\frac{d^2 a}{dt^2} 
\right]
\tilde{\beta} (t) =0.\nonumber
\end{eqnarray}

In high momentum situation,$p\gg ma,\ \ \frac{da}{adt}$,
the WKB solution takes the following form.
\begin{eqnarray}
\alpha^{(L)} &\sim&
\frac{1}{\sqrt{(2\pi)^3 2p }} \frac{p}{ma}
\left(
1-\frac{i}{pa}\frac{da}{dt} +O(p^{-2})
\right)
e^{-i\int dt\left[
p+\frac{m^2 a^2}{2p} -\frac{1}{2pa}\frac{d^2 a }{dt^2}
\right]}\nonumber\\
\beta^{(L)} &\sim& \frac{1}{\sqrt{(2\pi)^3 2p }}
\frac{p}{ma}(1+O(p^{-2}))e^{
-i\int dt\left[
p+\frac{m^2 a^2}{2p} -\frac{1}{2pa}\frac{d^2 a }{dt^2}
\right]
}.
\end{eqnarray}
Thus the longitudinal component is estimated in the high momentum limit
 as 
\begin{eqnarray}
A^{(L)}_\mu \sim  \frac{1}{\sqrt{(2\pi)^3 2p }}
\left[
i\partial_\mu \Lambda
-
\delta_{0\mu} \frac{ma}{p}
\exp
\left(
i\vec{p}\cdot\vec{x} -i
\int dt\left(
p+\frac{m^2 a^2}{2p} -\frac{1}{2pa}\frac{d^2 a }{dt^2}
\right)
\right]
\right),
\end{eqnarray}
where
$$
\Lambda =
\frac{1}{ma}
e^{i\vec{p}\cdot\vec{x} -i\int dt\left[
p+\frac{m^2 a^2}{2p} -\frac{1}{2pa}\frac{d^2 a }{dt^2}
\right]}.
$$

These solutions will be used in section 5 and 6.

%%%%%%%%%%%%%%%%%%%%%%%%%%%%%%%%%%%%%%%%%%%%%%%%%%%%%%%%%%%%%%%%%%%

\begin{flushleft}
\bf{4.\ \ Yukawa Interaction in the Expanding Universe}
\end{flushleft}

In this section we shall discuss high energy limit of  transition
probabilities  via  Yukawa interaction in the four dimensional universe.
The interaction is expressed by 
\begin{eqnarray}
S_{Yukawa} &=& \lambda \int d^4 x \sqrt{-g} 
 \phi (\bar{\Psi}_1 \Psi_2 + \bar{\Psi}_2 \Psi_1 )\nonumber\\
&=&
\lambda \int d^4 x \tilde{\phi}
( \bar{\tilde{\Psi}}_1 \tilde{\Psi}_2 + \bar{\tilde{\Psi}}_2 \tilde{\Psi}_1 )
\end{eqnarray}
where $\Psi_1$($\Psi_2$) is a spinor field with mass $m_1$($m_2$) and 
$\phi$ is a scalar field with mass $\mu$. The tilded fields are 
rescaled ones in section 2.

This action generates several types of 
 processes. For a 
 $\Psi_1$ particle  with conformal momentum $\vec{p}$ and helicity 
$1/2$ to decay into a $\Psi_2$ particle with $\vec{q}$ 
and a $\phi$ particle with $\vec{k}$, the amplitude is written down as
follows.
\begin{eqnarray}
Amp_{\Psi_1} = -i\lambda \int d^4 x 
e^{i(\vec{p}-\vec{q}-\vec{k})\cdot\vec{x}} 
u^{out\ast}_{\vec{k}} (t)
\bar{\tilde{\Psi}}^{out}_2  (h_f ,\vec{q},t)
\tilde{\Psi}^{in}_1 (h_i =1,\vec{p},t) \label{777}
\end{eqnarray}
where $u^{out}_{\vec{k}} (t)$ is an out-mode function of $\tilde{\phi}$ and 
$\tilde{\Psi}^{in/out} $ is an in-/out- mode function of $\tilde{\Psi}$
 and $h_f /2$ is helicity of the created spinor particle.
 In eqn(\ref{777}), we have omitted contribution from Bogoliubov 
 coefficients of vacuum polarization which is suppressed 
 in the high temperature limit .
In the following argument we assume that $m_1 <m_2 +\mu$, so no
decay happens in the flat spacetime limit. Even in the expanding 
spacetime the decay is prohibited especially 
 in  the past and future flat regions. Thus the decay is physically 
interpreted to occur only in the era of the cosmic expansion. 
 
 We expect that this assumption, $m_1 <m_2 +\mu$,  
is not essential for our results when  the decaytime we will obtain is
much shorter than that calculated ordinarily in the flat spacetimes, 
$E_1 /(m_1 \Gamma_{flat})$.

 In the expanding universe its scale factor is actually 
time dependent. Hence the meaning of probability {\it per unit time} seems
 ambiguous.  Therefore we adopt  transition probability
 itself to get a clear interpretation. The transition probability is
 defined by
\begin{eqnarray}
\sum |S|^2  = \int \frac{L^3 d^3 k}{(2\pi)^3}\frac{L^3 d^3 q}{(2\pi)^3}
\frac{1}{N^{out} N^{in} N_{\phi} }
|Amp_{\Psi_1}|^2 ,
\end{eqnarray}
where 
\begin{eqnarray}
N^{in/out} &=& \int_{L^3} dx^3 \ \bar{\tilde{\Psi}}^{in/out} \gamma^0 
\tilde{\Psi}^{in/out},\\ 
N_{\phi} &=& \int_{L^3} dx^3 i 
\left(
u_{\vec{k}}^{\ast} \partial_t u_{\vec{k}}
-
\partial_t u_{\vec{k}}^{\ast} u_{\vec{k}}
\right). 
\end{eqnarray}
Now our purpose is to analyze the high momentum limit
($p\rightarrow\infty$) 
of $\sum |S|^2$. As a warm up, let us first calculate a simple case
with the scale factor $a(t) = b\Theta (-t) +\Theta(t)$. It is  possible
to calculate the limit  straightforwardly. Substituting
 explicit form of mode functions into the definition of $\sum  |S|^2$,
we get
\begin{eqnarray}
&&W^{(step)}_{\Psi_1} (1/2\rightarrow h_f /2) 
=\lim_{p\rightarrow \infty} \sum |S|^2\nonumber\\
&=&\lim_{p\rightarrow\infty}
 \frac{\lambda^2}{64 \pi^3} \int d^3 q \nonumber\\
&&\times \left| \int^\infty_0 dt
\frac{e^{-it\left(\sqrt{p^2 + m_1^2} -\sqrt{q^2 + m_2^2}
-\sqrt{(\vec{p}-\vec{q})^2 + \mu^2}\right)}   }
{[(p^2 + m_1^2 )(q^2 + m_2^2)((\vec{p}-\vec{q})^2 + \mu^2 )]^{1/4} }
 \bar{U}_2 (h_f , \vec{q}, a=1) U_1 ( 1, \vec{p}, a=1)\right.\nonumber\\
&&\left.+
 \int^0_{-\infty}  dt
\frac{e^{-it\left(\sqrt{p^2 + m_1^2 b^2} -\sqrt{q^2 + m_2^2 b^2}
-\sqrt{(\vec{p}-\vec{q})^2 + \mu^2 b^2}\right)}   }
{[(p^2 + m_1^2 b^2)(q^2 + m_2^2 b^2)((\vec{p}-\vec{q})^2 + \mu^2 b^2)]^{1/4} }
 \bar{U}_2 (h_f , \vec{q}, a=b) U_1 ( 1, \vec{p}, a=b) \right|^2 ,\nonumber
\end{eqnarray} 
 where we have used the fact that the reflection component of the wavefunction
vanishes in the limit. 
After some tedious manipulation 
final analytic forms of $W$ turns out to be as follows. 
\begin{eqnarray}
W^{(step)}_{\Psi_1}
(1/2\rightarrow h_f =-1/2)&=&\frac{\lambda^2}{32\pi^2}
\left[
\frac{1+b^2}{1-b^2}\ln \frac{1}{b^2} -2
\right], \label{4.3}\\
W^{(step)}_{\Psi_1} (1/2\rightarrow h_f =1/2)&=&\frac{\lambda^2}{8\pi^2}
\left[1+\frac{b\ln b^2}{1-b^2} \right] F(m_1 , m_2 , \mu) ,\label{4.4}
\end{eqnarray}
where 
\begin{eqnarray}
F(m_1 ,m_2 , \mu) =\int^1_0 dy
\frac{(1-y)(m_1 y + m_2 )^2 }
{\mu^2 y + m_2^2 (1-y) -m_1^2 y(1-y) }.
\end{eqnarray}

\vspace{1cm}

Next we try to obtain  $W$ 
for arbitrary  $a(t)$ except
$a(\infty)=1$ and $a(-\infty) =b$.
Our strategy comes from  the fact that approximate energy conservation
holds for high momentum reactions as seen below.  
Now we concentrate on  manipulating 
 contribution from the phase space region where $p\gg ma$ and $q$
 and $k=|\vec{p}-\vec{q}|$ are of  order of $p$. 
As a result of this restriction 
 the WKB amplitude can be justified
 for both initial and final mode functions. 
 Consequently  $W$ is turned out to possess a factor like
\begin{eqnarray}
\Delta(h_f)&=&
\int^{\infty}_{-\infty} dt \bar{U}_2 (h_f ,\vec{q}, a(t))
U_1 (1, \vec{p},a(t)) \label{44.1}\\
&&\times\frac{
\exp
\left[
i\int dt
\left(
\sqrt{q^2 + m^2_2 a(t)^2}+\sqrt{(\vec{p}-\vec{q})^2 +\mu^2 a(t)^2 }
-\sqrt{p^2 + m^2_1 a(t)^2}
\right)
\right]
}
{\left[
 (p^2 +m^2_1 a(t)^2 )
 (q^2 +m^2_2 a(t)^2 )
 ((\vec{p}-\vec{q})^2 +\mu^2 a(t)^2 )
\right]^{1/4}} . \nonumber
\end{eqnarray}
It is very suggestive for $h_f =1$ 
to roughly estimate it by neglecting masses as follows.
\begin{eqnarray}
\Delta(h_f =1)&\sim& f(\vec{p}, \vec{q}) \times 
\int dt  e^{it(q+|\vec{p}-\vec{q}|-p)}  \nonumber\\
&=&f(\vec{p},\vec{q})\times (2\pi)^3 \delta (q+|\vec{p}-\vec{q}|-p).
\end{eqnarray} 
Therefore the following relation must be satisfied 
at least to this leading order. 
\begin{eqnarray}
p\sim q+|\vec{p}-\vec{q}|= q +\sqrt{(p-q)^2 +2pq(1-\cos \theta)}
\end{eqnarray} 
where $\vec{p}\cdot\vec{q} = pq\cos \theta$. This is approximate energy
conservation law and an important clue for us to calculate 
$W_{\Psi_1}$. From this ``conservation'' law
 only phase space region where $0< q< p$ and $\theta\sim0$ hold
 can contribute to $W_{\Psi_1}$. This fact tempts us to introduce 
a tiny but not specified constant $\theta_o \ll 1$ and 
a small constant $\epsilon$ satisfying 
$m_1 ,\ m_2 ,\ \mu \ll p\epsilon \ll p$. 
 Now  consider only a portion of the
phase space with $p\epsilon \leq q \leq p(1-\epsilon)$
 and $0\leq\theta\leq \theta_o $.
 Consequently  the following expansions are valid;
\begin{eqnarray}
&&\sin\theta\sim \theta ,\nonumber\\
&&1-\cos\theta \sim\frac{1}{2} \theta^2 .\nonumber
\end{eqnarray}
Furthermore as a result of the high momentum limit, we can expand 
\begin{eqnarray}
&&\sqrt{p^2 + m_1^2 a^2}\sim p + \frac{m_1^2 a^2}{2p},\\
&&\sqrt{q^2 + m_2^2 a^2}\sim q + \frac{m_2^2 a^2}{2q},\\
&&\sqrt{(\vec{p}-\vec{q})^2 + \mu^2 a^2}\sim 
p-q + \frac{pq\theta^2 + \mu^2 a^2}{2(p-q)} .
\end{eqnarray}
~From these useful expressions, we obtain for arbitrary $a(t)$ %%%%%%%%%!!!!!!!
\begin{eqnarray}
\lim_{p\rightarrow \infty} \sum |S(h_f =1)|^2 = W_{\Psi_1}(h_f=1) ,\nonumber
\end{eqnarray}
\begin{eqnarray}
W_{\Psi_1}
&=&\lim_{p\rightarrow \infty}
 \frac{\lambda^2}{32 \pi^3} \int^{p(1-\epsilon)}_{p\epsilon }
 d q \frac{q^2}{(p-q)}
\int^{\theta_o}_0 d\theta \theta^3 \nonumber\\
&&\times \left| \int^\infty_{-\infty} dt
e^{-i\int^t_0 dt(\frac{ m_1^2 a(t)^2}{2p} -\frac{m_2^2 a(t)^2}{2q}
-\frac{pq\theta^2 + \mu^2 a(t)^2}{2(p-q)} )}    
\right|^2\nonumber .
\end{eqnarray} 

Next change the integral variables as follows.
\begin{eqnarray}
q&=&py,\\
\theta &=& \frac{m_1}{p} z,\\
t &=& \frac{2p}{m_1^2} \eta .
\end{eqnarray}
Then $p$ dependence appears only in 
the upper bound of $z$ and the argument of
the scale factor. In fact the result is expressed such that
\begin{eqnarray}
W_{\Psi_1}(1/2\rightarrow
1/2)
&=&\lim_{p\rightarrow \infty}
 \frac{\lambda^2 m^2_1}{8 \pi^2} \int^{1-\epsilon}_\epsilon d y \frac{y^2}{1-y}
\int^{\theta_o \frac{p}{m_1}}_0 dz z^3 \label{44.123}\\
&&\times
\left|
\int^\infty_{-\infty} d\eta
\exp
\left[ 
\frac{1}{i}
\int^\eta_0 d\eta 
\left(
a^2
 -\frac{m_2^2 a^2}{m_1^2 y}
-\frac{ m_1^2 y z^2 + \mu^2 a^2 }
{m_1^2 (1-y)}
\right)
\right] 
 \right|^2 .\nonumber
\end{eqnarray}

To evaluate $W_{\Psi_1}$  the following relation is worth proving.  
\begin{eqnarray}
\lim_{p\rightarrow\infty} a=
\lim_{p\rightarrow\infty}
a\left( \frac{2p\eta}{m_1^2} \right)= b\Theta(-\eta) + \Theta (\eta) 
\label{44.10} .
\end{eqnarray}
Actually in the limit $\frac{2p\eta}{m_1^2}$ approaches  
$\infty$ when $\eta > 0$, while $-\infty$ when $\eta<0$.  From the fact
 that $a(\infty)=1$ and $a(-\infty)=b$ one can derive
easily the eqn (\ref{44.10}). By substituting (\ref{44.10})
 into (\ref{44.123}) we can proceed with calculation of  $W_{\Psi_1}$. 

Because we take $p\rightarrow \infty$, 
 $\theta_o p/m$ can be replaced into $\infty$. Hence no 
dependence of $\theta_o$ remains in the final form of $W_{\Psi_1}$. 
Moreover we 
can take  $\epsilon \rightarrow 0$ because of no appearance of infra-red 
divergence in the integral.   These
replacements yield
\begin{eqnarray}
&&W_{\Psi_1}(1/2 \rightarrow 1/2 ) \nonumber\\
&=&
 \frac{\lambda^2 m^2_1}{8 \pi^2} \int^1_0 d y (1-y)
\int^{\infty}_0 dz z^3 \nonumber\\
&&\times 
\left[ 
\frac{1}{z^2 + A(m_1; m_2 ,\mu , y)}
- 
\frac{1}{z^2 + b^2 A(m_1 ;m_2 ,\mu ,y) }
 \right]^2\nonumber ,
\end{eqnarray}
where
\begin{eqnarray}
A(m_1; m_2 ,\mu ,y)=
\frac{1}{y^2}
\left[\frac{m_2^2}{m_1^2} (1-y) +\frac{\mu^2}{m_1^2} y
-y(1-y) 
\right].
\end{eqnarray}
After integrations with respect to  $y$ and $z$, 
 the right hand side ends up to the same form of $W^{(step)}$.
\begin{eqnarray}
W_{\Psi_1}(h_f =1)=W^{(step)}_{\Psi_1} (h_f =1).\label{44.15}
\end{eqnarray}
It is a prominent feature  
 that this relation (\ref{44.15}) holds for arbitrary $a(t)$.   
This property implies the existence 
 of  notable universality 
 of $W_{\Psi_1} (h_f =1)$.

\vspace{1cm}

For $h_f = -1$ the story changes a little bit.
 The delta factor
 in eqn(\ref{44.1}) can be estimated roughly as 
\begin{eqnarray}
\Delta(h_f =-1)&\sim& f'(\vec{p}, \vec{q}) \times 
\int dt a(t) e^{it(q+|\vec{p}-\vec{q}|-p)}\\
&\sim& f'(\vec{p}, \vec{q}) F_1 (\Delta E ),
\end{eqnarray} 
where 
$\Delta E =q+|\vec{p}-\vec{q}|-p$ and
$$
F_1 (\Delta E) =\int dt a(t) e^{it\Delta E}.
$$
This scale factor contribution comes from  
\begin{eqnarray}
\bar{U}_2 (-1 , \vec{q}, a(t)) U_1 ( 1, \vec{p}, a(t))
\sim
\sqrt{pq}\left(\frac{m_1 }{p} +\frac{m_2 }{q} \right) a(t).
\end{eqnarray}

If $0<q<p$ and $\theta\ll 1$ do not  hold,  
    $\Delta E\sim O(p) \gg \omega^{(1)}_{max}$ is inevitably satisfied. 
Therefore,  as mentioned in the section 2,  $F_1 \sim 0$ holds. 
 Thus approximate energy conservation becomes valid again.
\begin{eqnarray}
&&0<q<p,\\
&&\theta\sim 0.
\end{eqnarray}
In this case it should be noticed that
  $\theta$ is not so large that $\Delta E$ becomes larger than 
$\omega^{(1)}_{max}$. Only phase space region with
\begin{eqnarray}
\theta< \theta_o =\sqrt{\frac{\omega^{(1)}_{max}}{p}}\label{4.2}
\end{eqnarray}
 contributes to $W_{\Psi_1}$. 
Estimation of $W_{\Psi_1}$ for $h_f =-1$ is also possible
 taking account of eqn(\ref{4.2}). 
Thanks to the high momentum limit the scale factor 
can be replaced the step one as $a(t)= b\Theta(-t) +\Theta (t)$.   
After manipulation  we can show 
$$W_{\Psi_1} (h_f=-1)=W^{(step)}_{\Psi_1} (h_f= -1).$$

\vspace{1cm}

Consequently  eqns (\ref{4.3}) and (\ref{4.4}) are turned out be 
 correct for arbitrary evolution of $a(t)$.  
 Eqn(\ref{44.10})  is the keypoint giving birth of the universality.
However in the actual universe $p$ does not take  infinite value but 
 finite and is of order of the temperature. 
It should be noticed for large but finite value of $p$  
that the replacement of $a(t)$ into the step 
evolution is valid only when a high momentum condition like 
\begin{equation}
\omega_{min} >  \frac{m^2}{p} \label{4.444}
\end{equation}
is satisfied. Here $\omega_{min}$ denotes 
the lowest typical frequency of $a(t)$ and is assumed of  
 the same order of minimum value of the Hubble parameter. The condition 
(\ref{4.444}) enables us 
to regard $a(t)$ as $b\Theta(-t) + \Theta(t)$.
 The inequality (\ref{4.444}) can be also rewritten as 
\begin{equation}
\frac{1}{m} > \left[\frac{m}{p}\right] \frac{1}{\omega_{min}},\label{44.11}
\end{equation}
where $1/m$ represents Compton length and $1/\omega_{min}$ means 
 maximum radius  of the Hubble horizon.  The factor $m/p\sim m/E$ can be 
interpreted as Lorentz contraction factor.  
 Imagine a particle with Compton length $1/m$ running with high momentum
$p$ in the expanding universe. 
 Also suppose that the universe begins to expand when the particle reaches 
a point A and that the universe ceases to expand when the particle
arrives at a point B. Length between A and B can be naively considered
 of order of the maximum Hubble horizon $(\sim1/\omega_{min})$.
Then the particle can get excitation energy from the
 gravitational field only while running between A and B. Meanwhile getting 
 $p$ larger, the length
 between A and B becomes Lorentz contracted as 
$[m/p] (1/\omega_{min})$ from the particle's viewpoint. Thus 
 when relation in eqn(\ref{44.11}) holds 
 the particle cannot see details of the way how the universe has
expanded. Consequently this yields the above universality.

The universality is actually very powerful tool 
to estimate the high momentum limit of the transition probability.
It can be shown that the universality also appear in  other kinds of 
interactions and in any dimension of the spacetime. 
However it should be noted that we must take  some care of 
 its treatment when  the probability grows up to infinity as
 $p\rightarrow\infty$. Using the universality 
 we can naively manipulate the limit explicitly. However, it might
diverge. In the later section such a phenomenon can be observed explicitly 
and then we should cut $p$ off at a certain large value of order of 
the temperature of the universe.

\vspace{1cm}

It is a rather straightforward application to
 estimate  $\phi$ particle decay probability 
into a $\Psi_1$ particle with helicity $h_1$ and a $\Psi_2$ particle
 with helicity $h_2$.  
Here we must assume that 
$\mu < m_1 + m_2$  in order to suppress the transition in the flat spacetime. 
The final forms of the transition probability are listed as follows.
\begin{eqnarray}
W_\phi (h_1 =1 , h_2 = 1)
=\frac{\lambda^2}{16\pi^2}
\left[
\frac{1+b^2}{1-b^2}\ln \frac{1}{b^2} -2
\right],\label{4.5}
\end{eqnarray}
\begin{eqnarray}
W_\phi (h_1 =1, h_2 =-1) =
\frac{\lambda^2}{8\pi^2}\left(
1+\frac{b}{1-b^2}\ln b^2
\right)
G(\mu,m_1 ,m_2 ) ,\label{4.6}
\end{eqnarray}
where 
\begin{eqnarray}
G(\mu ,m_1 ,m_2 )
=\int^1_0 dy \frac{m_1^2 (1-y)^2 -2m_1 m_2 y(1-y) +m_2^2 y^2}
{m^2_1 (1-y) +m_2^2  y -\mu^2 y(1-y)}.
\end{eqnarray}

\vspace{1cm}

Astonishingly it is found that all of these forms of the probabilities
 possess
 a certain kind of dual symmetry.  
Interchanging $b$ into $1/b$ keeps exactly the forms
 in eqns (\ref{4.3}), (\ref{4.4}), (\ref{4.5}), (\ref{4.6}).
This means that physics of 
the expanding universe is related with that of the contracting universe.
Note that the symmetry is not merely time reversal one 
because initial one particle decays two particles for both cases.    
 This might be related with a kind of hidden duality.

\vspace{1cm}

Finally we discuss decay rate of the particles. 
For example let consider decay of $\Psi_1$ particle. The decay  is
 expected to occur when $W_{\Psi_1}\sim 1$. For small $b$,
$W_{\Psi_1}$ behaves just as 
\begin{eqnarray}
W_{\Psi_1}
\sim\frac{N^{\ast} \lambda^2}{32\pi^2} \ln\frac{1}{b^2},
\end{eqnarray}
where $N^{\ast}$ is the number of final modes which 
 contribute to the decay.
Thus when the universe expands so enough that
\begin{eqnarray}
\frac{a_f}{a_i}=\frac{1}{b}\sim e^{\frac{16\pi^2}{N^{\ast}\lambda^2} }
\end{eqnarray}
 is satisfied, the particle gets decayed.
 Assuming the radiation dominant
universe, $b$ can be as follows.
\begin{eqnarray}
b
=\frac{a_i}{a_f}
=\sqrt{\frac{\tau_i}{\tau_f} }.
\end{eqnarray}
Consequently 
we obtain a decay rate;
\begin{eqnarray}
\Gamma_f
=\frac{1}{\tau_f}\sim 2
e^{-\frac{32\pi^2}
{N^{\ast}\lambda^2}}
H_i 
\end{eqnarray}
where $H_i =1/2\tau_i$ is the Hubble parameter at production time of
 the $\Psi_1$ particle.

%%%%%%%%%%%%%%%%%%%%%%%%%%%%%%%%%%%%%%%%%%%%%%%%%%%%%%%%%%%%%%%%%%%%%

\begin{flushleft}
\bf{5.\ \ Decay due to Three Point Vertex \\
\ \ \ \ in Arbitrary Dimension }
\end{flushleft}

 The geometric bremsstrahlung process is naturally expected to occur
 in arbitrary dimensional spacetimes. Furthermore  
the existence is not considered to depend on
 whether couplings of the reaction are dimensionful or not. Here we
give a rather simple example teaching us this feature. Let us
consider the $\phi^3$ theory with mass $m$ 
 in the n-dimensional spacetimes. The action reads
\begin{eqnarray}
S =
\int d^n x \sqrt{|g|} 
\left[
\frac{1}{2} \left(\nabla \phi \right)^2 -\frac{1}{2}\left( m^2 
-\frac{n-2}{4(n-1)} R
\right)
\phi^2
-\frac{1}{3!}\lambda \Lambda^{3-n/2}\phi^3
\right] ,
\end{eqnarray}
where we have introduced a mass parameter $\Lambda$. Hence the coupling 
constant $\lambda$ is dimensionless. 
 In $n\leq 6$ the vertex is merely a
renormalizable interaction.
 Meanwhile in $n>6$, the interaction is 
not renormalizable and then $\Lambda$ means some
cutoff scale of the theory.
Transition probability of
 a particle decay with conformal momentum $p$
 is straightforwardly given in the tree level as 
\begin{eqnarray}
\sum |S|^2=
(2\pi)^{2(n-1)} \lambda^2 \Lambda^{6-n} \int d^{n-1} q
\left|
\int dt a^{3-n/2} u^{\ast}_q u^{\ast}_{|\vec{p}-\vec{q}| } u_p
\right|^2\nonumber .
\end{eqnarray}
It is turned out soon that this expression itself converges 
for arbitrary fixed $p$. However taking $p\rightarrow \infty$,
 this 
 grows up to infinity for $n> 6$ . This comes from the fact that
 the vertex is nonrenormalizable for $n>6$. 
However there exists no trouble in the real universe because $p$ does
not take infinite value but that of order the universe temperature. 
Thus in the following argument
we treat $p$ finite but large compared with $m$ and the Hubble parameter. 
 Thus the probability
 is also finite. 
 To get a useful lower bound of the probability in 
$p \gg m$, 
we again pick up only contribution  from the phase space region with
 $0\leq\theta\leq\theta_o \ll 1$ and 
$p\epsilon \leq q \leq p(1-\epsilon)$.   
This implies that  
\begin{eqnarray}
\sum|S|^2 \geq W
&=&
(2\pi)^{2(n-1)} \lambda^2 \Lambda^{6-n} \frac{2 \pi^{n/2-1} }
{\Gamma \left(n/2 -1\right) }\nonumber\\
&&\times
\int^{p(1-\epsilon)}_{p\epsilon} dq q^{n-2} \int^{\theta_o}_0 
d\theta \sin^{n-3}\theta
\left|
\int dt a^{3-n/2} u^{\ast}_q u^{\ast}_{|\vec{p}-\vec{q}| } u_p
\right|^2\nonumber.
\end{eqnarray}
For $n\leq 6$ other phase space
 contributions vanishes in the high momentum limit and $W$ gives 
 us the probability itself, not merely a lower bound. 

Here assuming that $p\epsilon \gg m$, we can use the WKB wavefunctions
for both in- and out- mode functions. 
Replacing the integral
variables like
\begin{eqnarray}
q&=& py\\
\theta &=& \frac{m}{p} z\\
t&=& \frac{2p}{m^2} \eta ,
\end{eqnarray}
we obtain
\begin{eqnarray}
W
&=&
 \lambda^2 \left(\frac{m}{\Lambda}\right)^{n-6} \frac{ \pi^{n/2-1} }
{(2\pi)^{n-1}\Gamma \left(n/2 -1\right) }
\int^{1-\epsilon}_\epsilon dy \frac{y^{n-3} }{1-y} 
\int^{\theta_o \frac{p}{m}}_0 dz z^{n-3}\nonumber\\
&&\times
\left|
\int d\eta a^{3-n/2} 
e^{
i\int d\eta \left(
a^2\frac{1}{y}+\frac{y z^2 + a^2}{1-y} -a^2
\right)
}
\right|^2\nonumber.
\end{eqnarray}
By imposing $m^2 /p \ll \omega^{(3-n/2)}_{min}$  the replacement 
 $a= b\Theta(-\eta)+\Theta(\eta)$
 is again validated. 
Then  the integration with respect to $\eta$ can be easily 
performed and we obtain
\begin{eqnarray}
W
&=&
 \lambda^2 
\left(
\frac{m}{\Lambda}
\right)^{n-6} 
\frac{ \pi^{n/2-1} }
{(2\pi)^{n-1}\Gamma \left(n/2 -1\right) }
\int^{1-\epsilon}_\epsilon dy y^{n-5}(1-y)\nonumber\\
&&\times \int^{\theta_o \frac{p}{m}}_0
 dz z^{n-3}\left[
\frac{1}{ z^2+C}
-\frac{b^{3-n/2} }
{ z^2 + b^2 C }
\right]^2 ,\label{5.61}
\end{eqnarray}
where $C=A(m; m,m,y)=(1-y+y^2)/ y^2$.

Unfortunately for $n=2$ this expression 
is not suitable and we must calculate
separately. However the estimation is also possible analytically 
and it results in just  
\begin{eqnarray}
W(n=2) =0.\nonumber
\end{eqnarray}
~From $n=3$ to $n=6$ we can take $p\rightarrow\infty$ and $\epsilon
\rightarrow 0$ in eqn (\ref{5.61}). The results are as follows. 
\begin{eqnarray}
&&W(n=3)=
\frac{\lambda^2}{8\pi}\left(\frac{4}{3}-\ln 3  \right)
\left(\frac{\Lambda}{m}\right)^3 
\left[
1 -\frac{2\sqrt{b}}{1+b}
\right].\nonumber\\
&&W(n=4)=
\frac{\lambda^2}{8\pi^2}
\left(
\frac{4}{\sqrt{3}}\arctan
\left(
\frac{1}{\sqrt{3}}
 \right)
-1
\right)
\left(\frac{\Lambda}{m}\right)^2 
\left[
1 +\frac{b}{1-b^2} \ln b^2 
\right].\nonumber\\
&&W(n=5)=
\frac{\lambda^2}{32\pi^2}\left(\frac{5}{4}\ln 3 -1  \right)
\frac{\Lambda}{m}
\left[
1 -\frac{2\sqrt{b}}{1+b}
\right].\nonumber\\
&&W(n=6) =
\frac{\lambda^2}{384\pi^3}
\left[
\frac{1+b^2}{1-b^2}\ln \frac{1}{b^2} -2
\right].\nonumber
\end{eqnarray}
Note that all these expressions are invariant 
 under a transformation $b\rightarrow 1/b$.

As mentioned previously $W$ diverges for $n>6$ in $p\rightarrow\infty$. 
Keeping $\omega^{(3-n/2)}_{min} > m^2 /p$ in mind, a lower
 bound with $\theta_o =m/p$ is obtained from eqn (\ref{5.61}). 
The explicit form with $b\sim0$ is  as follows.
\begin{eqnarray}
W(n>6, b\sim 0)
 \geq
\frac{ \pi^{n/2-1} }
{(2\pi)^{n-1}\Gamma \left(n/2 -1\right) }
\frac{\lambda^2}{(n-3)(n-4)(n-6)}
\left(
\frac{m}{\Lambda b}
\right)^{n-6} . \nonumber
\end{eqnarray}
Therefore after when the universe expands enough to satisfy   
$$
1/b\sim O\left(\lambda^{-\frac{2}{n-6}} \frac{\Lambda}{m}\right), 
$$
the particle gets decayed via the geometric bremsstrahlung.

\vspace{1cm}

After all  we arrive at a conclusion that in spite of the high
 temperature of the universe
  the transition probability due to the Yukawa geometric bremsstrahlung does
 not vanish  even in the n-dimensional spacetimes except $n=2$.

%%%%%%%%%%%%%%%%%%%%%%%%%%%%%%%%%%%%%%%%%%%%%%%%%%%%%%%%%%%%%%%%%%%%%%%%

\begin{flushleft}
\bf{6.\ \ Decay Process including Massive Gauge Field}
\end{flushleft}

In this section we shall survey decay processes including 
a massive vector particle $A_\mu$ with mass $\mu$. 

Let us first consider
that this particle interact with fermions $\Psi_1$ and $\Psi_2$ 
with mass $m_1$ and $m_2$. The interaction term can be expressed as
\begin{eqnarray}
S_{\Psi\Psi A}&=&g \int d^4 x  \sqrt{-g}
 \bar{\Psi}_1(c_V + c_A \gamma^5 ) \gamma^\mu \Psi_2 A_{\mu} + c.c \nonumber\\
&=& g\int d^4 x  \bar{\tilde{\Psi}_1}
(c_V + c_A \gamma^5 )
\gamma^a \tilde{\Psi}_2 \tilde{A}_a
+c.c 
\end{eqnarray}
where $A_\mu =e_\mu^a A_a =a\delta^a_\mu A_a$ and $ A_a =a^{-1} \tilde{A}_a$.
In the high momentum limit we can use again the WKB
approximation in the section 2. 
This enables us to calculate $W$ explicitly.

Let us consider first decay of a transverse component of $A_\mu$
possessing conformal momentum $p$ into 
$\Psi_1$ and $\bar{\Psi}_2$ particles. Just as the Yukawa interaction case,
we only take account of contribution from restricted phase space region
 as follows.
\begin{eqnarray}
&&W_{T} \left(h,\bar{h}\right)\nonumber\\
&=&\lim_{p\rightarrow\infty}
\frac{g^2}{32 \pi^2} \int^{p(1-\epsilon)}_{p\epsilon} dk \frac{k} {p(p-k)} 
\int^{\theta_o}_0
d\theta \theta \nonumber\\
&&\times \left|
\int dt 
\bar{U}_1 (h, \vec{p}-\vec{k},a)(c_V + c_A \gamma^5 )
 \gamma^\mu V_2 (\bar{h} , \vec{k}, a) 
\epsilon_\mu^{(T)}(\vec{p}, a)
e^{
-i\int dt 
\left(
\frac{\mu^2 a^2}{2p} 
-
\frac{pk\theta^2 +m^2_1 a^2 }{2(p-k)}
-\frac{m^2_2 a^2}{2k}
\right)
}
\right|^2 \nonumber,
\end{eqnarray}
where 
\begin{eqnarray}
\epsilon^{(\pm)}_\mu (\vec{p},a)
=\pm \frac{1}{\sqrt{2}} [ 0, 1, \pm i,0 ].
\end{eqnarray}
For any combination of $(h,\bar{h})$, it can be straightforwardly 
 shown that $W_T$ converges
 in $p\rightarrow\infty$ and $\epsilon \rightarrow 0$. 
After  a straightforward manipulation we obtain
 the following results.
\begin{eqnarray}
W_T (h=1 , \bar{h}= 1)
=
\frac{g^2}{4\pi^2}
\left(
1+\frac{b}{1-b^2}\ln b^2
\right)
H(m_1 ,m_2 ,\mu)
,\nonumber
\end{eqnarray}
where
$$
H =\int^1_0 dy
\frac
{|c_V +c_A |^2 m^2_1 y^2 + |c_V -c_A |^2 m^2_2 (1-y)^2 
+2(|c_V |^2 -|c_A |^2 ) m_1 m_2 y(1-y)}
{m^2_1 y + m^2_2 (1-y) -\mu^2 y(1-y)} .
$$

\begin{eqnarray}
W_T (h =-1,\bar{h} =-1)
=0.
\end{eqnarray}

\begin{eqnarray}
W_T (h =1,\bar{h} =-1)
=
\frac{g^2}{24\pi^2} | c_V -c_A |^2
\left[
\frac{1+b^2}{1-b^2}\ln \frac{1}{b^2} -2
\right] .\nonumber
\end{eqnarray}

\begin{eqnarray}
W_T (h =-1,\bar{h} =1)
=
\frac{g^2}{24\pi^2} | c_V + c_A |^2
\left[
\frac{1+b^2}{1-b^2}\ln \frac{1}{b^2} -2
\right] .\nonumber
\end{eqnarray}

To get W for longitudinal component decay is achieved by 
 taking the photon helicity vector as 
\begin{eqnarray}
\epsilon^{(L)}_\mu 
\sim
\frac{k}{\mu a} [ 1, 0, 0, -1 ].
\end{eqnarray}
After some manipulation  $W_L$ is in fact converged
 with $h=1$ and $\bar{h}=-1$  as
\begin{eqnarray}
W_{L} (h=1, \bar{h} =-1)
=\frac{g^2}{8\pi^2}
\left[
1+ \frac{b\ln b^2}{1-b^2} 
\right] K(m_1 ,m_2 ,\mu, c_V , c_A )
,\nonumber
\end{eqnarray}
where
\begin{eqnarray}
&&K(m_1 ,m_2 , \mu, c_V ,c_A )\nonumber\\
&=& \int^1_0 dy \frac{y(1-y)}{m_1^2 y +m_2^2 (1-y) -\mu^2 y(1-y) }
\nonumber\\
&&\times
\left|
\frac{c_V (m_1 -m_2) -c_A (m_1 +m_2)}{\mu} 
m_1 \sqrt{\frac{y}{1-y}} \right.\nonumber\\
&&\ \ \ -
\frac{c_V (m_1 -m_2) +c_A (m_1 +m_2)}{\mu} 
m_2 \sqrt{\frac{1-y}{y}} \nonumber\\
&&\ \ \ \left. -2(c_V +c_A )\mu \sqrt{y(1-y)}
\right|^2 .
\end{eqnarray}

\begin{eqnarray}
W_{L} (h=-1, \bar{h} =1)
=\frac{g^2}{8\pi^2}
\left[
1+ \frac{b\ln b^2}{1-b^2} 
\right] K(m_1 ,m_2 ,\mu, c_V , -c_A ).
\end{eqnarray}

While $W(h=\pm 1, \bar{h}= \pm 1)$ is given as 
\begin{eqnarray}
W_L (h=\pm 1,\bar{h} =\pm 1 )
&=&
\frac{g^2}{16\pi^2}
\frac{|c_V (m_1 -m_2 ) \pm c_A (m_1 + m_2 )|^2}{\mu^2}
\left[
\frac{1+b^2}{1-b^2} \ln \frac{1}{b^2} -2
\right]
.\nonumber
\end{eqnarray}

\vspace{1cm}

It is also possible to estimate $W$ when  
a $\Psi_1$ particle with momentum $p$ and helicity $1/2$ decays into 
 a $\Psi_2$ particle and a $A_\mu$
particle. Substituting the wavefunction forms in the section 2 into $W$ 
 it is written down explicitly
 for emission of transverse component such that
\begin{eqnarray}
&&W_{\Psi_1} \left(h_i ;h_f,T \right)\nonumber\\
&=&\lim_{p\rightarrow\infty}
\frac{g^2}{32 \pi^2} \int^{p(1-\epsilon)}_{p\epsilon} dk \frac{k} {p(p-k)} 
\int^{\theta_o}_0
d\theta \theta \nonumber\\
&&\times \left|
\int dt 
\bar{U}_2(h_f, \vec{p}-\vec{k},a)
(c_V + c_A \gamma^5 ) \gamma^\mu U_1 (h_i , \vec{p}, a) 
\epsilon_\mu^{\ast (T)}(\vec{k}, a)
e^{
i\int dt \Delta E
}
\right|^2 ,\nonumber
\end{eqnarray}
where 
$$
\Delta E =
\frac{\mu^2 a^2}{2k} 
+
\frac{pk\theta^2 +m^2_2 a^2 }{2(p-k)}
-\frac{m^2_1 a^2}{2p}.
$$

For cases with $h_i =1$ and $h_f =1$ our WKB results include
 an infra divergence. 
\begin{eqnarray}
W_\Psi \left(1;1, 1\right)
=
\frac{g^2}{8\pi^2} |c_V -c_A|^2
\left(
\frac{1+b^2}{1-b^2}\ln \frac{1}{b^2}-2
\right)
\int^1_\epsilon \frac{dy}{y} 
.\label{5.2}
\end{eqnarray}

\begin{eqnarray}
W_\Psi \left(1;1, -1\right)
=
\frac{g^2}{8\pi^2} |c_V -c_A|^2
\left(
\frac{1+b^2}{1-b^2}\ln \frac{1}{b^2}-2
\right)
\int^1_\epsilon \frac{dy}{y} (1-y)^2
.\label{5.22}
\end{eqnarray}

In spite of the appearance in eqns (\ref{5.2}) and (\ref{5.22})  
of infra-red divergence with respect to $p\epsilon$,
 we believe that exact form of the probability converges as a result of the
existence of natural infra-red cut off $\mu$. This cut off works only in the 
 low momentum region where the WKB approximation is invalid. Because
we have discussed only a case where the WKB approximation is valid,
 our treatment is supposed to include such 
a superficial infra-red divergence. From this
point of view, it can be  naively expected that $\epsilon \sim \mu/p$.

However if $\mu$ is exactly zero, no natural infra-red cut off appear.
This infra-red divergence  really exists  
owing to the existence of the soft particle. The
number of  soft particles 
is counted only to the accuracy of observations. 
Thus for very soft particles one cannot discriminate  between
  virtual emission which contributes particle mass operator and real emission. 
 This enables us to add a part of the mass operator contribution to 
 the emission probability. Then the theory is hoped to have
 no real infra-red divergence in  physical interpretation,  
similarly to the flat spacetime case\cite{BN}.

For other cases of transverse emission, $W$ is converged and 
final results are given by
\begin{eqnarray}
W_{\Psi_1} (1;-1,1)=\frac{g^2}{4\pi^2}
\left[
1+\frac{b}{1-b^2}\ln b^2
\right]
R(m_1 , m_2 ,\mu) ,
\end{eqnarray}
where
$$
R= \int^1_0 dy\frac{y(
|c_V -c_A |^2 m^2_2 -2 (|c_V |^2 -|c_A |^2)
m_1 m_2 (1-y) +|c_V + c_A |^2 m_1^2 (1-y)^2 ) }
{ m^2_2 y +\mu^2 (1-y) -m_1^2 y(1-y) }.
$$
$$
W_{\Psi_1} (1;-1,-1) =0 .
$$

\vspace{1cm}

For the emission of longitudinal component we get
\begin{eqnarray}
&&W_{\Psi_1} \left(h_i ;h_f,L \right)\nonumber\\
&=&\lim_{p\rightarrow\infty}
\frac{g^2}{32 \pi^2} \int^{p(1-\epsilon)}_{p\epsilon} dk \frac{k} {p(p-k)} 
\int^{\theta_o}_0
d\theta \theta \nonumber\\
&&\times \left|
\int dt 
\bar{U}_2 (h_f, \vec{p}-\vec{k},a) (c_V + c_A \gamma^5 )
\gamma^\mu U_1 (h_i , \vec{p}, a) 
\epsilon_\mu^{\ast (L)}(\vec{k}, a)
e^{
i\int dt (\Delta E +\frac{1}{2k} \left( \frac{da}{adt} \right)^2
}
\right|^2 .\nonumber
\end{eqnarray}
Note that  $W_{\Psi_1} (1:1,L)$ 
possesses an infra-red divergence
 with respect to $\epsilon \rightarrow 0$.
\begin{eqnarray}
&&W_{\Psi_1} (1;1,L)\nonumber\\
&&=\frac{g^2}{8\pi^2}
\left(
1+\frac{b\ln b^2}{1-b^2}
\right)
Q(m_1 ,m_2 ,\mu)
 ,\nonumber
\end{eqnarray}
where 
\begin{eqnarray}
Q&=&\int^1_\epsilon dy
\frac
{y(1-y)}
{
\mu^2 (1-y) +m^2_2 y -m^2_1 y(1-y)
}\nonumber\\
&&\times
\left|
\frac{c_V (m_1 -m_2 ) +c_A (m_1 +m_2)}{\mu} 
\frac{m_2}{\sqrt{1-y}} \right.\nonumber\\
&&\ \ \ +
\frac{c_V (m_1 -m_2 ) -c_A (m_1 +m_2)}{\mu} 
 m_1 \sqrt{1-y} \nonumber\\
&&\ \ \ \left. -2(c_V -c_A ) \mu \frac{\sqrt{1-y}}{y}
\right|^2 
\end{eqnarray}
and it is expected that $\epsilon \sim \mu /p$.

Finally it can be shown that 
 $W_{\Psi} (1;-1,L)$ converges and takes the following form.
$$
W_\Psi (1;-1,L)=\frac{g^2}{32 \pi^2}\left| 
\frac{c_V (m_1 -m_2 ) -c_A (m_1 +m_2)}{\mu}
\right|^2
\left[
\frac{1+b^2}{1-b^2} \ln \frac{1}{b^2} -2
\right].
$$

Let us next comment on a three point interacting between  the massive 
 vector field with mass $\mu$ and  complex scalar fields $\varphi_1$
 and $\varphi_2$ with mass $m_1$ and $m_2$ .
The interaction is described by 
$$
S_{\bar{\varphi}\varphi A}
=
g \int d^4 x \sqrt{-g} A_{\mu} i
\left(
\bar{\varphi}_1 \nabla^{\mu} \varphi_2 -\nabla^{\mu} \bar{\varphi}_1 
\varphi_2
\right) +c.c. .
$$ 
We can also give in this case explicit forms of $W$ in each process 
as follows.

For decay of the transverse component ,
$$
W^{(T)}=\frac{g^2}{48\pi^2} 
\left[
\frac{1+b^2}{1-b^2} \ln\frac{1}{b^2} -2
\right].
$$

For the longitudinal component,
\begin{eqnarray}
W^{(L)}=&&\frac{g^2}{8\pi^2 \mu^2}
\left(
1+\frac{b\ln b^2}{1-b^2} 
\right)\nonumber\\
&&\times
\int^{1/2}_{-1/2} dx (1 -4 x^2)
\frac{(m^2_1 -m^2_2)^2 +4\mu^2 (m^2_1 -m^2_2 ) +4\mu^4 x^2}
{2m^2_1 +2m^2_2 -\mu^2  +4(m^2_1 -m^2_2) x 
+ 4\mu^2 x^2 }.\nonumber
\end{eqnarray}

For transverse vector particle emission from a  scalar particle,
$$
W_{\varphi_1}^{(T)}=\frac{g^2}{8\pi^2}\ln\frac{1}{\delta}
\left[
\frac{1+b^2}{1-b^2} \ln\frac{1}{b^2} -2
\right].
$$

For longitudinal vector particle emission out of a scalar particle,
\begin{eqnarray}
W_{\varphi_1}^{(L)}&=&\frac{g^2}{8\pi^2 \mu^2}
\left(
1+\frac{b\ln b^2}{1-b^2} 
\right)
\int^1_\delta \frac{dy}{y} 
\frac{(1-y)[(m^2_1 -m^2_2 )y +(2-y)\mu^2 ]^2 }
{\mu^2 (1-y) +m^2_2 y -m^2_1 y(1-y)}\nonumber\\
&\sim&
\frac{g^2}{2\pi^2 }\ln \frac{1}{\delta}
\left(
1+\frac{b\ln b^2}{1-b^2} 
\right).\nonumber
\end{eqnarray}

Thus probability of geometric bremsstrahlung does not vanish, 
 even including the vector field.

%%%%%%%%%%%%%%%%%%%%%%%%%%%%%%%%%%%%%%%%%%%%%%%%%%%%%%%%%%%%%%%%%

\begin{flushleft}
\bf{Acknowledgment}
\end{flushleft}

We would like to thank T. Futamase for valuable discussion.
Also we wish to thank R. Blankenbecler and S. Drell for fruitful 
 discussion about multiple scattering effect to geometric bremsstrahlung 
in highly densed matter.

The work of M.H. is supported  in part by the Grant-in-Aid for 
Science Research from the Ministry of Education
(No.08740185).

%%%%%%%%%%%%%%%%%%%%%%%%%%%%%%%%%%%%%%%%%%%%%%%%%%%%%%%%%%%%%%%%


\begin{thebibliography}{100}

\bibitem{FHIY}
T. Futamase, M. Hotta, H. Inoue and M. Yamaguchi,\\
{\it Geometric bremsstrahlung in the Early Universe},\\
to appear in Prog.Theor.Phys. , Vol 96, No1 (1996).


\bibitem{KT}
One can see a standard 
review of the early universe from the viewpoint of the 
high energy particle physics, for example, in \\
E.W. Kolb and M.S. Turner, {\it The Early Universe}(Addison-Wesley,
1990).


\bibitem{BD}
N.D.Birrell, Ph.D Thesis, King's College, London, (1979).\\
N.D.Birrell and J.G.Taylor, {\it J.Math.Phys.}(N.Y.){\bf
21},1740(1980).\\
N.D.Birrell and P.C.W.Davies, {\it Quantum Fields in Curved Space}
(Cambridge University Press, 1982).


\bibitem{BN}
F. Bloch and A. Nordsieck, {\it Phys.Rev.}{\bf 52}, 54(1937).




\end{thebibliography}
\end{document}